\documentclass[aps,final,notitlepage,oneside,twocolumn,nobibnotes,nofootinbib,superscriptaddress,noshowpacs,centertags]{revtex4-1}
\usepackage[english]{babel}
\usepackage{graphicx}
\usepackage{latexsym}
\usepackage{amssymb}
\usepackage{amsmath}
\usepackage{float}

\begin{document}

\title{Cosmological horizons as they are looked from a moving frame}
\author{V. I. Dokuchaev}\thanks{e-mail: dokuchaev@inr.ac.ru}
\affiliation{Institute for Nuclear Research of the Russian Academy of Sciences, Moscow, Russia}
\affiliation{National Research Nuclear University MEPhI (Moscow Engineering Physics Institute), Moscow, Russia}
\author{Yu. N. Eroshenko}\thanks{e-mail: eroshenko@inr.ac.ru}
\affiliation{Institute for Nuclear Research of the Russian Academy of Sciences, Moscow, Russia} 

\date{\today}

\begin{abstract}
We consider the cosmological horizons in the expanding universe from the point of view of observer moving with respect to CMB frame. The deformation (non-sphericity) of cosmological horizons is demonstrated. Some principle consequences are discussed.
\end{abstract}

\maketitle

In cosmology, one refers to as the ``system at rest'' the system in which the CMB has no dipole anisotropy. In
relation to such system the Sun moves with velocity 370~km~s$^{-1}$ towards constellation of
the Lion. This preferential frame $K$ was called ``C-frame'' in the works \cite{SatTat72}.
Another definition of this frame based on the isotropy of Hubble flow, this is also the frame
in which the  baryon matter is locally at rest. In differential geometry C-frame lies at the
hypersurface, perpendicular to the lines of time. Probably, this frame was selected at the
epoch of reheating after inflation due to spontaneous symmetry breaking onto the particular
quantum vacuum state \cite{SatTat72}.

Let us consider the two systems of coordinates: system $K$ which is at rest with respect to CMB
and system K', moving freely  with respect to $K$ in the positive direction of axes X, and we
suppose that at the moment $t=t_0$ the origins of the systems $K$ and $K'$ coincide. We use the
operational definition of cosmological horizon (particle horizon) proposed by Rindler as ``a
frontier between things observable and things unobservable'' \cite{Rin56}. The particle
horizon, i.e. the maximum causally connected region, in $K$ is a sphere of the radius $R_{\rm
H}=a(t_0)\int_0^{t_0}dt/a(t)$, where $a(t)$ is the scale factor of the universe with metrics
\begin{equation}
ds^2=dt^2-a^2(t)(dx^2+dy^2+dz^2), \label{frw}
\end{equation}
and we restrict ourselves only by the flat space. The function $a(t)$ depends on the details of
cosmology. 

\begin{figure}[t]
\begin{center}
\includegraphics[angle=0,width=0.49\textwidth]{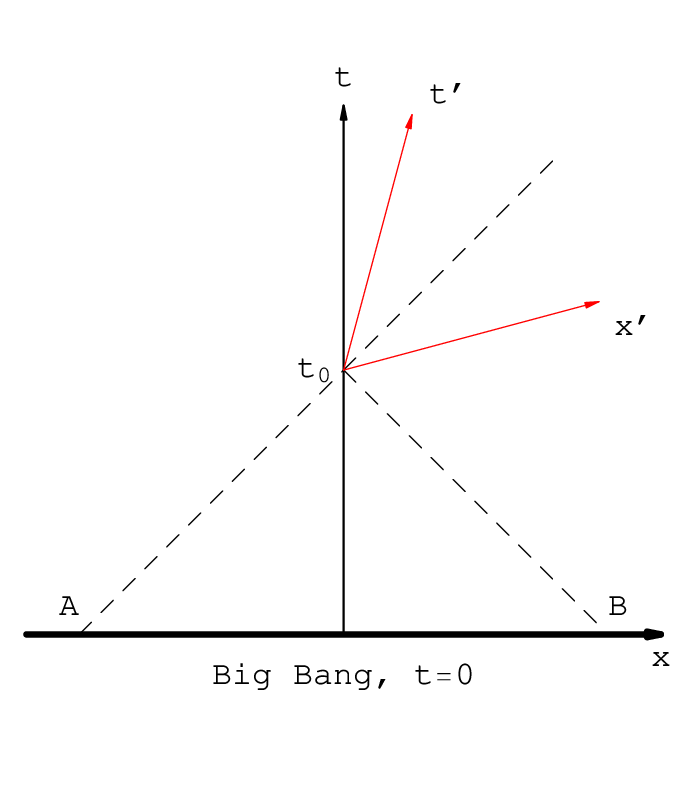}
\end{center}
\caption{Schematic Minkowsky diagram for the static toy cosmological model. Segment AB is the particle horizon of the co-moving observer $O$ at $t=t_0$, $x=0$. The arrows show the coordinate axes of the observer $O'$, having Lorentz boost at the moment $t_0$. For the boosted
observer the size of AB is larger in comparison with the measurements of the
co-moving observer O.} \label{mind}
\end{figure}

Imagine for a moment that we are in Minkovski space which appeared at $t=0$ by the clock of the
system $K$, see Fig.~\ref{mind}. In this toy model the observer in $K$ at $t=t_0$ has the spherical horizon with
radius $R_H=ct_0$. The points of the hypersurface $t=0$ are not simultaneous in the moving
system $K'$. They have different $t'$ at different points in $X'$ direction, as it is obvious
from the Lorentz transformations. Let us consider the converging spherical light wave that
focused at the common beginnings of $K$ and $K'$ at the moment $t=t'=t_0$ and put that this
wave was emitted near the moment $t=0$. For the observer in the system $K$ the emitters are on
the spherical surface S with radius $R_{\rm H}$. For the observer in $K'$ the corresponding
surface S' represents a surface of ellipsoid with semiaxes $\gamma R_{\rm H}$, $R_{\rm H}$ and
$R_{\rm H}$, where $\gamma$ - is the Lorentz-factor. Note that the surface S' is elongated (not
shortened for a moving observer!) because its boundary is the surface of simultaneity in the
system $K$ (see also \cite{Lev04}). As the converging light front in the both systems remains
spherical, it is clear that in the system $K'$ atoms of the surface S' have radiated light not
simultaneously, and atoms in $+X'$ direction radiated first. Here the preference of the system
$K$ among all over inertial systems arises due to our choice of $t=0$ as the time birth by the
clocks of $K$. The universe's expansion somehow complicates these arguments, but the effect of
horizon deformation is roughly the same.

Now we return to the real universe. It's difficult to construct the rigid system $K'$ in the
whole expanding universe. Instead, we define the flexible moving system by the following way.
Let us consider the set of test (dust) particles all moving freely in the positive direction of
axes X with equal velocities $u^\mu=dx^\mu/ds$. Their physical components
$U^\mu=a(t)u^\mu\propto1/a(t)$, $U^1=U$, $U^2=U^3=0$ and the ordinary velocity is expressed as
$v^2=U^2/(1+U^2)$. Let us consider some event in $K'$ and the test particle, which coincides
with this event. We put the time of the event equals to the proper time of the particles passed
from the $K$ and $K'$ coincidences at $t=t_0$. So we calculate the interval along the world
line of the test particle:
\begin{equation}
d\tau^2=\left.ds^2\right|_{\mbox{~along the w.l.}}=\frac{dt^2}{1+U^2}, \label{tautr}
\end{equation}
therefore
\begin{equation}
\tau=\int\limits_{t_0}^t dt'\left[1+U^2(t_0)\frac{a(t_0)}{a(t')}\right]^{-1/2}+const. \label{tautrint}
\end{equation}

As the space coordinate we choose the coordinate of the test particle under consideration in
$K$-system at $t=t_0$:
\begin{equation}
x'=x-\int\limits_{t_0}^{t}\frac{Udt}{\sqrt{1+U^2}}, \label{transf1}
\end{equation}
and the transverse coordinates are $y'=y$ and $z'=z$. These space coordinates are therefore the
Lagrange coordinates of the test particles. In the system $K'$ with the above coordinates transformations the interval
takes the form
\begin{equation}
ds^2=d\tau^2-2aUd\tau dx'-a^2(dx'^2+dy'^2+dz'^2). \label{metrnew}
\end{equation}
In the metrics (\ref{metrnew}) the ideal fluid energy density is transformed as
\begin{equation}
\varepsilon'=T_{0'0'}=\gamma^2(\varepsilon+pv^2), \label{vareps}
\end{equation}
which coincides with the local Lorentz transformation law \cite{LanLif-II} (\S~35), and it's
interesting that the phantom matter gets the negative energy density.

\begin{figure}[t]
\begin{center}
\includegraphics[angle=0,width=0.49\textwidth]{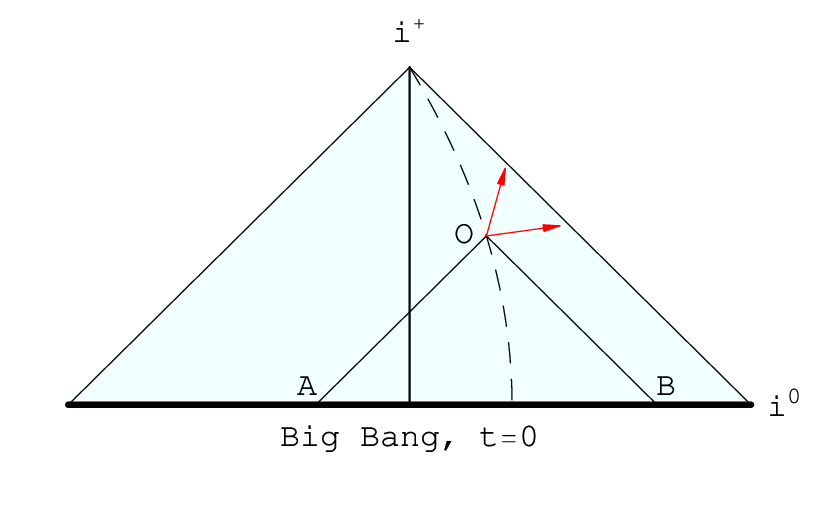}
\end{center}
\caption{Schematic Carter-Penrose diagram of Freedman-Robertson-Walker universe in the $t-x$
plane. Segment AB is the particle horizon of the co-moving observer $O$ at $t=t_0$. The arrows
show the tetrad of the observer $O'$, having Lorentz boost at the moment $t_0$. The boosted
observer $O'$ sees the larger physical size of AB then the co-moving observer O does.} \label{pend}
\end{figure}

The light cone obeys the differential equation $ds^2=0$. In the Y' and Z' directions the light
cone and the horizon size are the same as in the original $K$-system. In the X' direction the
light front moves according to the expressions
\begin{equation}
x'_\pm=\int\limits_0^{\tau}d\tau (-U\pm\gamma)/a(t)+C_\pm=\int\limits_{t_0}^{t}dt(-v\pm1)/a(t),
\label{rayeq}
\end{equation}
where the signs $\pm$ mark the two direction of the light propagation, and the constants of
integration $C_\pm$ in the last expression were fixed by the conditions $x'_\pm(t=t_0)=0$. The
full coordinate size of the particle horizon in X' direction (distance between the events A and
B at Fig.~\ref{pend}) is
\begin{equation}
\Delta x'=x'_{-}(0)-x'_{+}(0)=2\int\limits_0^{t_0}dt/a(t)=2R_H(t_0)/a(t_0).
\end{equation}
For the future event horizon the analogous expression reads as
\begin{eqnarray}
\Delta x'_F&=&x'_{+}(t=\infty)-x'_{-}(t=\infty)= \\
&=&2\int\limits_{t_0}^{\infty}dt/a(t)=2R_F(t_0)/a(t_0). \nonumber 
\end{eqnarray}

Therefore the coordinate sizes of the horizons are the same in $K$ and $K'$: $\Delta x'=\Delta
x$, $\Delta y'=\Delta y$ and $\Delta z'=\Delta z$. But the physical size in $K'$ is equal to
\cite{LanLif-II} (Chapter X, \S 84)
\begin{eqnarray}
dl'^2&=&\left(-g_{\alpha\beta}+\frac{g_{0\alpha}g_{0\beta}}{g_{00}}\right)dx^{\alpha}dx^{\beta}= \nonumber \\
&=&a^2(\gamma^2dx'^2+dy'^2+dz'^2),
\label{len}
\end{eqnarray}
where $\gamma(t)=(1-v^2(t)/c^2)^{-1/2}$. The expression (\ref{len}) taken at $\tau=const$ means
that the horizons sizes (both particle and future event horizons) are larger in $K'$-system in X'
direction in $\gamma(t)$ times in comparison with $K$-system. Schematic Carter-Penrose diagram of Freedman-Robertson-Walker universe in the $t-x$
plane is shown at Fig.~\ref{pend}.

An observer in $K'$ sees the bulk motion of all the matter (except for the test particles),
which is at rest in $K$. This situation is similar to the anisotropic cosmological models with
bulk motion instability \cite{ZelNov75} (Chapter 19, \S 5).

One may consider an observer that originally at rest with respect to CMB frame K, accelerated
for some time interval $\delta t$, and then moves freely with $K'$. From the point of view of
this observer the causal region elongated from sphere to ellipsoid. Note again that the nonsphericity of the causally-connected area in $K'$ is the result of the existence of preferred system $K$
(C-frame) connected with CMB in the expanding universe.

To avoid confusion with the problem of the apparent shape of a relativistically moving sphere,
we must note that the horizon deformation under consideration is defined by causality reasons,
in contrast with aberration effect \cite{Pen59}, which is purely local phenomenon. Therefore
the discussion between \cite{Lev04} and \cite{CalGomMotReb04} is not applicable to our case, in
particularly because we consider an observer {\it inside} the surface of a horizon.

The observational quantitative consequences of the horizon deformation are very small. But from the principle point of view we can outline the two possible effects. First, the deformation leads to the Lorentz noninvariance of holographic dark energy models \cite{Li04}. This is because the UV cutoff in the holographic approach obeys the local Lorentz invariance while the IR cutoff does not. The later cutoff is settled by the cosmological horizon which deforms for the moving observer. Second, for the calculation of the cosmological Casimir energy \cite{Zeld68} the boundary conditions in $K$ and in the moving $K'$ are different. This is because the  Compton length of the virtual particle
can not exceed the maximum causal scale, i.e. the horizon scale. This effect can change the Casimir energy density by a tiny amount.

The reported study was partially supported by RFBR, research project No. 15-02-05038~a.

\end{document}